\documentclass[12pt]{iopart}
\usepackage{graphicx}

\begin{document}

\title[A new general purpose event horizon finder]{A new general purpose
event horizon finder for 3D numerical spacetimes}

\author{Peter Diener}

\address{Max-Planck-Institut f\"{u}r Gravitationsphysik, 
Albert-Einstein-Institut, Am M\"{u}hlenberg 1, D-14476 Golm, Germany}

\ead{diener@aei.mpg.de}

\begin{abstract}
I present a new general purpose event horizon finder for full 3D
numerical spacetimes. It works by evolving a complete null surface backwards
in time. The null surface is described as the zero level set of a scalar
function, that in principle is defined everywhere. This description of the
surface allows the surface, trivially, to change topology, making this
event horizon finder able to handle numerical spacetimes, where two (or
more) black holes merge into a single final black hole.
\end{abstract}

\pacs{04.25.Dm, 04.70.Bw, 95.30.Sf, 97.60.Lf}

\submitto{\CQG}

\section{Introduction}
An event horizon (EH) is defined as a 2+1 surface in 3+1 space, inside of
which no null geodesics can reach future null infinity, while outside at least
some can. For that reason an EH is a global concept and can in 
principle only be found when the full history of the spacetime is known. 
The horizon itself is generated by outgoing null geodesics, that once they
have joined onto the horizon will forever stay on it.

In recent years 3D numerical evolutions of binary black hole spacetimes
have become stable enough that it is possible in some cases to follow the 
final merged black hole for a significant time \cite{Alcubierre02a}.
Therefore it makes sense to start looking for EHs in these 
numerical spacetimes. However, previously published EH finders for
numerical spacetimes \cite{Hughes94a, Anninos94f, Libson94a, Masso98c} have
either been too slow, limited to single black hole spacetimes or have taken
advantage of special symmetries in order to handle changes in topology and
can not be used in the general 3D case without symmetries.

This paper is organized in the following way. In section~\ref{sec:basic} I
will present the basic ideas and methods of EH finding. In 
section~\ref{sec:surface} I will discuss different ways of describing the
surface and present the level set description used here. The numerical
implementation is described in section~\ref{sec:numerics}, where also a cure
for a very serious problem is presented.  In section~\ref{sec:test_analytic}
I will present tests of the EH finder using analytical data with
known location of the EH to test the accuracy of the code while in
 section~\ref{sec:test_numeric}
I will present tests using numerical data in order to test the robustness of
the code. Finally in section~\ref{sec:discuss} I will discuss prospects of
future work and uses of the code.
\section{Event horizon finding}
\label{sec:basic}
In principle the EH can be found by integrating null geodesics forward in time.
Outgoing null geodesics just outside or inside the EH will all diverge away 
from the EH. Those outside will escape to infinity, while those inside will 
end up at the singularity. However, as pointed out in \cite{Libson94a} it is
practically impossible to follow the generators of the horizon forward in
time, since small numerical errors will cause the null geodesics to deviate
exponentially away from the EH. This can, on the other hand, be taken 
advantage of, if the generators are integrated backwards in time, since then 
the EH will be an attractor. A method based on this was presented in 
\cite{Anninos94f}.

For the backward null geodesic integration method, the presence of a small
tangential velocity component can cause the null geodesics to deviate from 
the EH. Even though the null geodesic may return to the EH, it will be in
a different position and it might even cross other null geodesics, thereby
creating spurious caustics. The cure is to evolve the complete horizon
surface as a whole. Since the only way a surface can move is in its normal
direction, tangential drift is not an issue. This can be done by representing
the surface by a function
\begin{equation}
f(t,x^{i}) = 0, \label{eq:f_define}
\end{equation}
and requiring this surface to satisfy the null condition
\begin{equation}
g^{\alpha\beta}\partial_{\alpha}f\partial_{\beta}f = 0. \label{eq:f_null}
\end{equation}
Expanding equation~\eref{eq:f_null} out yields a quadratic equation for
$\partial_{t}f$, which can be solved, giving the following evolution equation
for $f$
\begin{equation}
\partial_{t}f = \frac{-g^{ti}\partial_{i}f+\sqrt{(g^{ti}\partial_{i}f)^{2}-
g^{tt}g^{ij}\partial_{i}f\partial_{j}f}}{g^{tt}}. \label{eq:f_evolve}
\end{equation}
Here the root was chosen so as to describe outgoing null geodesics.
Notice that, in contrast to the geodesic equation, this equation does not
contain any derivatives of the metric.

Note that the exact location of the EH is unknown at the end of a numerical
evolution. However the location of the apparent horizon (AH) can serve as a
good initial guess for the EH (see \cite{Nakamura84, Tod91, Kemball91a,
Baumgarte96, Thornburg95, Gundlach97a, Libson94b, Alcubierre98b,
Shoemaker-Huq-Matzner-2000, Huq00, Schnetter02a, Thornburg2003:AH-finder} for
details about finding AHs). It will be completely inside the EH, but will
be very close if the numerical spacetime is almost stationary. A practical way
of locating the EH is to start with the AH as an initial guess for the EH
and to use the attracting property of the EH to approach it asymptotically.
It is therefore important to note that the surface $f=0$ will 
never exactly coincide with the EH, but will approach it exponentially. In 
order to estimate when the surface $f=0$ is a good approximation, a different 
initial surface (chosen so as to be completely outside of the EH) can be 
evolved backwards in time. The EH will then always be located between these 
two surfaces and when the two surfaces agree to within a small fraction of a 
grid cell the location of the EH is known with sufficient accuracy.

\section{Description of the surface}
\label{sec:surface}
In \cite{Libson94a} either of the following parametrization was adopted 
to describe the surface in the presented axisymmetric cases
\begin{eqnarray}
f(t,r,\theta) & = & r-s(t,\theta), \\
f(t,z,\rho) & = & \rho-s(t,z), \label{eq:cylindrical}
\end{eqnarray}
where equation~\eref{eq:cylindrical} was used in the colliding black hole case, in 
order to allow the surface to cross itself thereby describing the locus of 
generators before they join onto the EH. This was possible due to the high
level of symmetry, but would be difficult to generalize to the non symmetric
case, since it requires prior knowledge of the location of the caustic 
points (in this case the symmetry axis).

The choice made in this work, is to avoid any parametrization of the surface
that might run into trouble, but to keep the description in 
equation~\eref{eq:f_define}. That is, the horizon surface is described as the 
0-level isosurface of the scalar function $f$, where $f$ is negative inside 
and positive outside the surface. There are several advantages to this choice.
The main advantage is that changes of topology (such as when black holes 
merge) are handled naturally with no special symmetry requirements. What 
happens in these cases is simply that the number of regions with negative 
$f$-values changes. Another advantage is that the function $f$ is defined 
at the same grid points as the numerical metric so that no interpolation is 
necessary in order to evolve the surface. The last point can also be seen as a 
disadvantage, since when $f$ is defined on fixed grid points only, there are 
in general no grid points exactly on the $f=0$ isosurface. This means that 
in order to analyze the surface it is necessary to find it first. A second
(and more serious) problem is the fact that during evolution of $f$ its
gradients steepen. This can be seen in the following way: The level set
function $f$ actually defines an infinity of surfaces with different
iso-values. Thus for example the triplet of iso-values $f=[-0.5,0.0,0.5]$ 
defines three outgoing null surfaces that will all move towards the EH, when
evolved backwards in time. This means that these three surfaces will move
towards each other and since they are defined by constant values of $f$, the
gradients will steepen. This will cause numerical problems if it is not
addressed. Another potential problem is illustrated in \ref{ap:kerrschild},
where it is shown that for at least one choice of the metric (Schwarzschild in
Kerr-Schild coordinates), the evolution of $f$ is ill conditioned.

\section{Numerics}
\label{sec:numerics}

The evolution of equation~\eref{eq:f_evolve} is performed using the Method of Lines
(MoL) with either second order Runge-Kutta (RK2) or 3-step Iterative Crank
Nicholson (ICN). The spatial derivatives of $f$ are calculated using one
sided second order derivatives. The directions of the derivatives are
determined by different schemes depending on the problem. If the shift is
zero, the directions are determined using the values of $f$ itself according
to the scheme in \ref{ap:upwind}. On the other hand if the shift is non-zero,
the direction of the shift has to be taken into account in order to preserve
stability, the main reason being that the evolution equation contains terms of
the form $g^{ti}\partial_{i}f$. In this case the stencil direction is taken to 
be opposite the shift direction, since I am evolving backwards in time. 

In order to be able to handle numerical data with excised regions, the code 
supports dynamic interior excision regions, that can be used even when the
numerical data is defined everywhere. Since the EH normally does
not occupy the whole numerical domain it is only necessary to output a
rectangular box that contains the EH at all times. This can result
in a very significant reduction of the amount of required disk space for the
numerical metric data. Furthermore to save computation time, once the numerical
data has been read in, the computationally active region can be further reduced
so that only a smaller rectangular box surrounding the black holes is used. As
the holes move across the grid, this smaller active region moves along with
them by de-activating and activating grid points as necessary. At the
interior boundaries and the boundaries of the rectangular box, one sided 
second order derivatives are used as well.

The initial guess for $f$ at $t=T$, where $T$ is the time at the end of the
numerical evolution, is normally chosen as a sphere of radius $r_{0}$ 
centered at the point $P=(x_{0},y_{0},z_{0})$
\begin{equation}
f(T,x,y,z) = \sqrt{(x-x_{0})^{2}+(y-y_{0})^{2}+(z-z_{0})^{2}}-r_{0},
\end{equation}
which has the nice property that
\begin{equation}
|\nabla f|=1, \label{eq:f_grad}
\end{equation}
except at the origin where it is non differentiable. The parameters for the
initial guess are chosen to be as good a guess for the EH as possible, while
making sure that $f=0$ is either completely inside or outside of the EH. If
the EH is judged (for example by looking at the shape of the AH) to be
significantly nonspherical it is also possible to use an arbitrarily 
oriented and shaped ellipsoid as the initial guess. This has not been
necessary so far since, with the currently used gauge conditions, the final
horizon is almost always close to being a coordinate sphere.

As mentioned, there is a problem with gradients of $f$ steepening during the
evolution. For that reason $f$ is re-initialized regularly during the 
evolution so that it again satisfies equation~\eref{eq:f_grad} approximately. To
obtain this re-initialization the evolution equation
\begin{equation}
\frac{\rmd f}{\rmd \lambda} = -\frac{f}{\sqrt{f^{2}+1}}
\left (|\nabla f|-1\right )
\label{eq:f_reinit}
\end{equation}
is evolved in an unphysical parameter $\lambda$ until a steady state has been
achieved. The factor in front of the parenthesis in equation~\eref{eq:f_reinit} 
consists of two parts. The numerator is present to make sure that no evolution takes
place when $f=0$, since I do not want to move the surface during
re-initialization. The denominator is there to make sure that the Courant
condition for stability is limited by the constant value, 1, instead of 
the maximum value of $f$. The term in the parenthesis makes sure that
the evolution stops when $|\nabla f|=1$. The evolution of 
equation~\eref{eq:f_reinit}
is also done with MoL, but since it is only the final steady state that are
significant, it is sufficient to use a simple Euler scheme. However, a second
order Runge-Kutta scheme has been implemented as well. The spatial derivatives
used for the re-initialization are second order upwinded according to the
scheme in \ref{ap:upwind}. Since the re-initialization equation does not
depend on the shift, the shift direction is ignored.

If the re-initialization is performed often enough, I do not have to worry
about potential stability problems due to the use of one sided differences
of $f$ at the boundaries. This is the main reason, that it is possible to
evolve only a small region around the $f=0$ surface. Note also, that without
re-initialization all different $f=const$ surfaces would be evolved as
null surfaces, but every time re-initialization is done all surfaces
except $f=0$ will be changed, so they can not any more be considered to be
null surfaces. In this way $f=0$ is picked out as the surface of interest.

The re-initialization equation~\eref{eq:f_reinit} is similar to the one used
in \cite{Sussman94}, the differences being introduced to make the surface
move as little as possible, while still being reasonably fast.

This re-initialization scheme works well most of the time, but it has some
problems when there is only a few points in either directions inside the
surface. In those circumstances the level 0 isosurface can move significantly
outwards. This typically occurs just before the topology of the surface
changes. To avoid this, a scheme has been implemented to detect when this
happens and avoid doing the re-initialization until after the topology
has changed. The detection scheme essentially consists of searching for
all places where 1) $f$ is negative (i.e.\ inside the surfaces) and 2) there
is a local extremum of $f$ in all directions. Since the value of $f$ in such a
grid point is approximately equal to the signed distance from that grid point
to the level 0 isosurface, the absolute value of the maximum of $f$ in these
points can be used to estimate the minimum width of the region with $f<0$. 
If this width is less than a given threshold, the re-initialization is
not done.

In order to calculate the areas of the EHs, it is necessary to
locate points on the surfaces. This task is complicated by the fact that in
principle, the number of individual surfaces at any given time is not known.
Therefore points on the surfaces present are located in several separate
steps. 

First the number of surfaces in the data is found. This is done using a 
``flooding'' algorithm using an integer mask ({\tt surface\_mask}) initialized
to zero at the grid points of the level set function, $f$, and an integer 
counter ({\tt surface\_counter}) initialized to zero. The algorithm proceeds
as follows. Locate a grid point where $f<0$ and {\tt surface\_mask=0}. If 
such a grid point does not exist there are no surfaces, so exit. Otherwise 
increment {\tt surface\_counter} and mark this point
{\tt surface\_mask=\tt surface\_counter}. Repeatedly find connected grid 
points inside the same surface, i.e.\ locate points with $f<0$ and at 
least one neighbour point with {\tt surface\_mask=\tt surface\_counter} and
mark those likewise until no more such points can be found. In this way all 
points inside the first surface has been marked. Now check if another point 
with $f<0$ and {\tt surface\_mask=0} exists. This point must then be located 
inside another surface. Repeat the above marking procedure until all points 
with $f<0$ has been marked with a surface number.

Next an approximate coordinate centroid is found for each surface using the
marked points. This is used as the center for a polar coordinate system 
used to explicitly parametrize the corresponding surface as $r(\theta,\phi)$.
For each chosen direction (given by $\theta$ and $\phi$) the radius of the
surface is found using interpolated values of $f$ in a Newton iteration root 
finder\footnote{A Hermite interpolation polynomial is used, since not only 
the function but also the derivative has to be continuous in the Newton 
iteration.}.

With this explicit parametrization of the surface, the area, centroid and
circumferences can be easily computed. Unfortunately it is not always possible
to find this explicit parametrization. As shown later in Section~\ref{sec:3bh}
the EH can in some cases be so distorted that with any choice
of center, $f$ would be multivalued along some angular directions. For this
reason a more general surface finding and integration routine using a direct
triangulation of the surfaces is under consideration.

I would like to emphasize that the use of interpolation is only necessary
when finding points on the surfaces for analysis purposes and is not used
at all in the evolution and re-initialization of the level set function.
  
The EH finding algorithm has been implemented as a thorn in
Cactus \cite{Allen99a, Goodale02a} using the Einstein Toolkit 
\cite{Goodale03c} and is fully parallelized. Though it might happen that the
computationally active region is contained completely on one processor, this
is usually not a serious concern since the thorn uses a lot less memory than
a comparably sized numerical evolution and can therefore be run on a lot less
processors. Also it is not necessary to run the EH finder on the
full computational grid of the spacetime evolution. It suffices to output just
the part of the grid that completely contains the EH (or horizons)
within it at all times, also cutting down on the required disk space for output.

The source code for the EH finder, will be released for public use in the
near future.
\section{Tests using analytical data}
\label{sec:test_analytic}
In this section I will present tests of the code using analytic spacetimes,
where the location of the EH is known analytically at all times.
\subsection{Non-rotating black hole in isotropic coordinates}
The first test uses the metric for a non-rotating black hole in
isotropic coordinates. In these coordinates the metric is
\[
\rmd s^{2} = -\left (\frac{1-M/(2r)}{1+M/(2r)}\right )^{2}\rmd t^{2}+
\left ( 1+\frac{M}{2r}\right )^{4}[\rmd r^{2}+r^{2}(\rmd \theta^{2}+
\sin^{2}\theta \rmd \phi^{2})],
\]
where $M$ is the mass of the black hole.
In these coordinates the horizon is a stationary sphere with radius
$r_{EH}=M/2$. The equation of motion for radial outgoing null geodesics is
\[
\frac{\rmd r}{\rmd t} = \frac{1-M/(2r)}{\left (1+M/(2r)\right )^3}.
\]
Looking at null geodesics close to the EH $r=M/2+\epsilon$ where
$\epsilon<<M/2$ and expanding it can be seen that
\[
\frac{\rmd \epsilon}{\rmd t} \approx \frac{\epsilon}{4M}.
\]
So integrating forward in time the null geodesic diverges exponentially from
the EH. However integrating backwards in time it should be
expected that the null surface converges exponentially to the location of the
EH with an e-folding time of $4M$.

\begin{figure}[ht]
\begin{center}
\includegraphics[width=150mm]{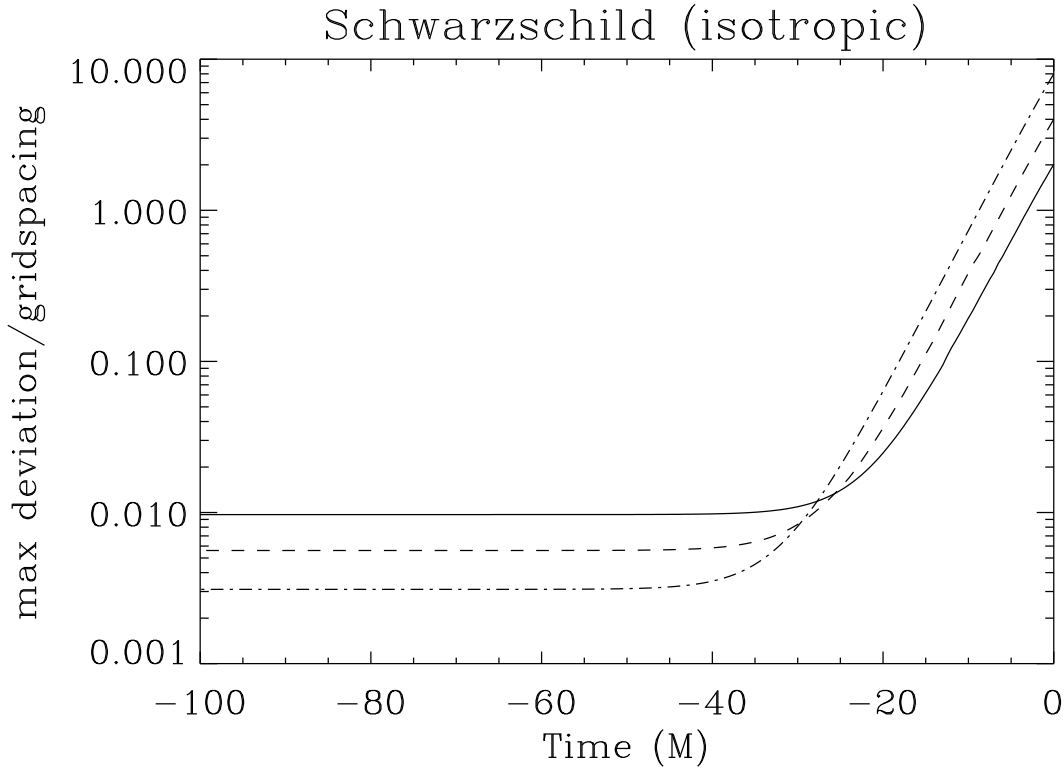}
\end{center}
\caption{The maximum deviation of the surface from the analytic EH
divided by the gridspacing for a Schwarzschild black hole of mass $M=1$ in 
isotropic coordinates as function of time. The solid line is $\Delta=0.05M$,
the dashed line is $\Delta=0.025M$ and the dash-dot line is $\Delta=0.0125M$.}
\label{fig:schw_err}
\end{figure}

I performed runs at three different resolutions $\Delta = 0.05M, 0.025M,
 0.0125M$ for $M=1$ with an initial radius for the null surface equal to 
$r_{0}=0.4M$ (i.e.\ 2, 4 and 8 gridspacings away from the true EH).
In figure~\ref{fig:schw_err} I show the maximum distance of the surface from
the true location of the EH divided by the gridspacing as a 
function of time in a logarithmic plot.
Initially there is a clear exponential convergence until the finite difference
solution is reached with a numerical e-folding time of $4.13M$, agreeing
nicely with the analytic expectation. At higher resolution the 
finite difference solution is closer to the analytic solution. At a resolution
of $\Delta = 0.05$ the maximum error at $T=-100M$ is $0.0097\Delta$. At the
higher resolution of $\Delta = 0.025$ the corresponding error is $0.0056\Delta$,and at the highest resolution of  $\Delta = 0.0125M$ the error is 
$0.0031\Delta$. Looking at the ratio of the error at successive resolutions, it
can be seen that the code is approaching second order convergence. Looking at
the absolute errors it is worth noting that the method is able to locate the 
EH to less than 1/100 of a grid spacing when the radius of the 
EH is only 10 grid spacings.

\subsection{Non-rotating black hole in Kerr-Schild coordinates}
The second test uses the metric for a non-rotating black hole in
Kerr-Schild coordinates. In these coordinates the metric is
\begin{equation}
\rmd s^{2}=\left (-1+\frac{2M}{r}\right ) \rmd t^{2} + 
\frac{4M}{r} \rmd t \rmd r + \left ( 1+\frac{2M}{r}\right ) \rmd r^{2} +
r^{2} (\rmd \theta^{2} + \sin^{2}\theta \rmd \phi^{2}).
\label{eq:schw_ks}
\end{equation}
Just like in the previous section it is easy to find the equation of 
motion for radial outgoing null geodesics and analyse the limit of null
geodesics, close to the EH (in these coordinates $r_{EH}=2M$).
The result is that the e-folding time is $4M$ for these coordinates as well.

\begin{figure}[ht]
\begin{center}
\includegraphics[width=150mm]{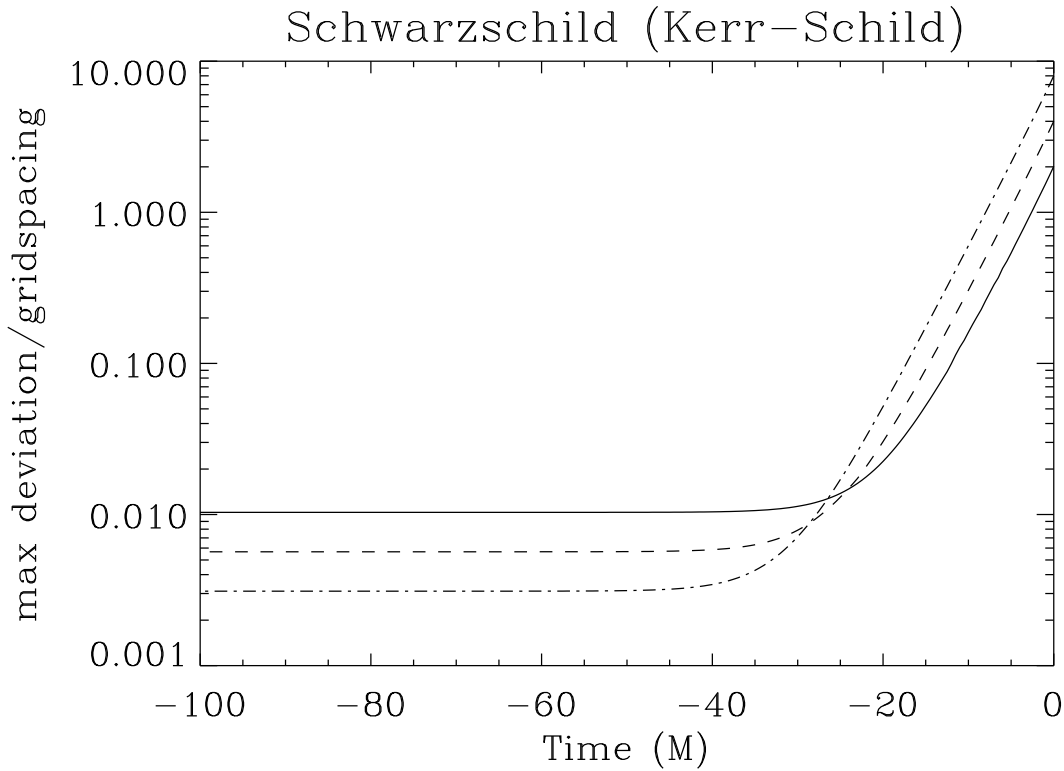}
\end{center}
\caption{The maximum deviation of the surface from the analytic EH
divided by the gridspacing for a Schwarzschild black hole of mass $M=1$ in
Kerr-Schild coordinates as function of time. The solid line is $\Delta=0.2M$, 
the dashed line is $\Delta=0.1M$ and the dash-dot line is $\Delta = 0.05M$.}
\label{fig:schw2_err}
\end{figure}

Here I performed runs at three different resolutions $\Delta = 0.2M, 0.1M,
0.05M$ for $M=1$ with an initial radius for the null surface equal to 
$r_{0} = 1.8M$ (i.e.\ 2, 4 and 8 gridspacings away from the true EH).
In figure~\ref{fig:schw2_err} I show the maximum distance of the surface from
the true location of the EH divided by the gridspacing as a
function of time in a logarithmic plot.
Again there is a clear exponential convergence until the finite difference
solution is reached with a numerical e-folding time of $4.01$ in excellent
agreement with the analytic expectation. Also in this case, the higher the
resolution the closer the finite difference solution is to the analytic
solution. The maximal error divided by the grid spacing at $T=-100M$ is
0.01, 0.0057, 0.0031 respectively at the 3 different resolutions used. I.e.\
very similar to the results for Schwarzschild in isotropic coordinates.

\subsection{Rotating black hole in Kerr-Schild coordinates}
The third test uses the metric for a rotating black hole in Kerr-Schild
coordinates. In these coordinates the EH is an ellipsoid with the
following equation
\begin{equation}
d\equiv\frac{x^2+y^2}{r_{+}^2+a^{2}}+\frac{z^{2}}{r_{+}^{2}}-1 = 0, 
\label{eq:deviation}
\end{equation}
where $r_{+}=M+\sqrt{M^{2}-a^{2}}$, $M$ is the mass and $a=J/M$ is the angular
momentum per unit mass. For points not located on the EH the
deviation, $d$, defined in equation~\eref{eq:deviation} can be used as
a measure of the error. However it will not be a direct measure of the
distance from the surface to the EH. Usually it will be somewhat
larger than the distance.

\begin{figure}[ht]
\begin{center}
\includegraphics[width=150mm]{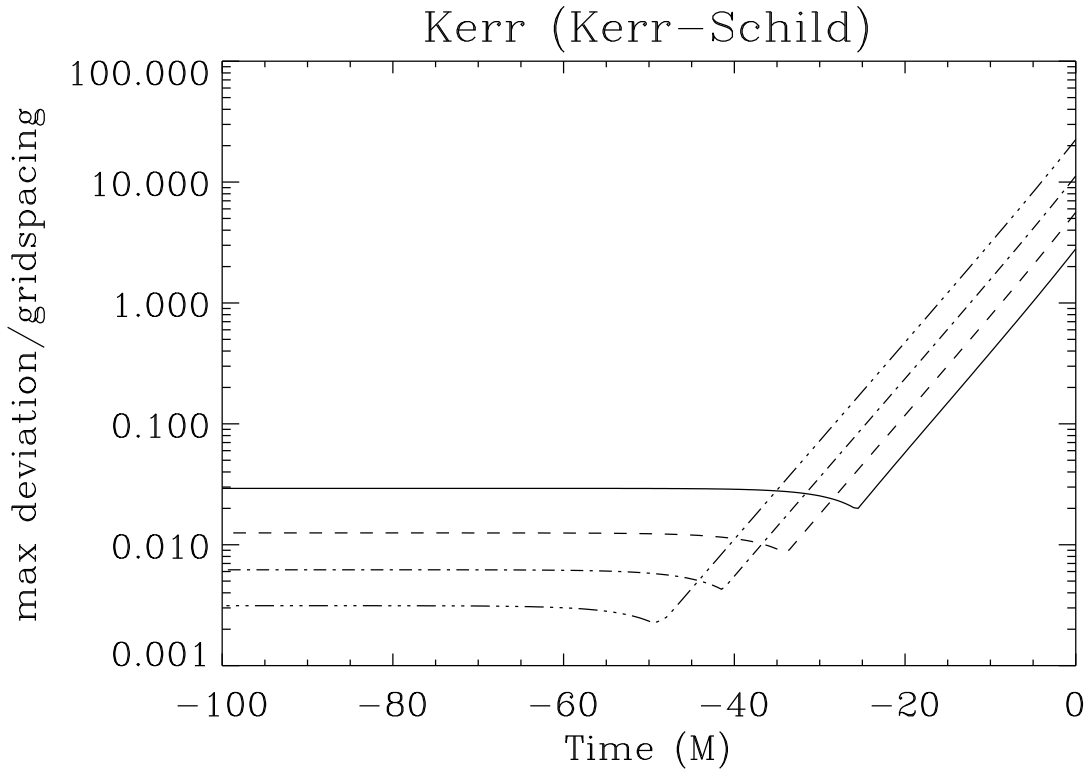}
\end{center}
\caption{The maximum deviation, as defined in equation~\eref{eq:deviation}, 
divided by the gridspacing for a Kerr black hole of mass $M=1$ and $a=0.8$ in
Kerr-Schild coordinates as function of time. The solid line is $\Delta=0.2M$,
the dashed line is $\Delta=0.1M$, the dash-dot line is $\Delta = 0.05M$ and 
the dash-triple-dot line is $\Delta = 0.025M$.}
\label{fig:ks_err}
\end{figure}

I performed runs at four different resolutions $\Delta = 0.2M, 0.1M,
0.05M, 0.025M$ for $M=1$ and $a=0.8$ with an initial radius for the null 
surface equal to $r_{0} = 2.0$ in all cases. In figure~\ref{fig:ks_err} I 
plot the deviation as defined in equation~\eref{eq:deviation} divided by the 
gridspacing as a function of time in a logarithmic plot.
At all resolutions there is a nice exponential convergence until the finite
difference solution is reached with a numerical e-folding time of $5.18$.
The deviation, $d$, divided by the grid spacing at $T=-100M$ goes down at higher
resolution. For the four resolutions it is $0.029, 0.013, 0.0062, 0.0031$
respectively, showing perfect second order convergence.

\section{Tests using numerical data}
\label{sec:test_numeric}
In this section I will present results from tests based on metric data from
highly dynamic spacetimes evolved numerically with Cactus. The exact location
of the EH is not known in these numerical spacetimes, but I compare to other
published results to the extent possible. At the same time it will be a test
of the robustness of the code and how well it can handle changes of topology
in multiple black hole spacetimes.
\subsection{Misner data with $\mu = 2.2$}
In~\cite{Misner60} initial data describing two black holes, initially at rest,
was presented. The initial separation between the two black holes is described
by a parameter $\mu$. These types of initial data have been extensively studied
in the literature and especially the EH for these data was studied using
numerical data from axisymmetric 2D codes and from 3D codes in 
\cite{Anninos94f}, \cite{Libson94a} and \cite{Anninos96c}.

The spacetime was evolved using the BSSN formulation of the Einstein
equations developed in \cite{Shibata95} and \cite{Baumgarte99} and implemented
in Cactus in \cite{Alcubierre99c}. In order to be able to compare with the 
results in \cite{Anninos94f} I used maximal slicing. The run was done at fairly
low resolution of $\Delta = 0.128M$ in order to be performed with a modest
amount of output on a workstation. With the recent development of improved
shift conditions in \cite{Alcubierre02a}, it turned out to be
advantageous to use the hyperbolic version of the gamma freezing shift. In
this way the slice stretching was controlled, the expansion of the
horizon in coordinate space was strongly reduced and it was possible to evolve
for a very long time. In the present case, the evolution was halted at $T=70M$,
since this is long enough to track the EH accurately for the first
$50M$ of the evolution.

At $T=70M$ the AH was found in the numerical data and it 
turned out to be nearly spherical with a coordinate radius of
$r_{AH}=2.84M$ (the ratio of the equatorial to the polar circumference was
$C_{eq}/C_{pol}=1.00074$). Two different runs with the EH finder were
therefore started at $T=70M$; one run where the initial surface described
by $f=0$ was a sphere of radius $r_{I}=2.5M$ completely contained within the
AH and another run with an initial sphere of radius $r_{O}=r_{I}+0.5M=3.0M$
which seemed to be a safe guess for a surface completely outside of the EH.
This was confirmed by the fact that initially the inner surface was
expanding while the outer surface was contracting when the surfaces were
evolved backwards in time.

Both of these surfaces were evolved successfully backwards in time, through
the change of topology all the way to the initial data slice. The change of
topology occurred at around $T\approx 2.75M$.

In \cite{Libson94a} it can be seen from  figure~11 that the topology changes
between $T=2M$ and $T=3.3M$. However, it turns out that a more precise
transition time is $T\approx 3.2M$ \cite{SeidelPrivateComm}, which is somewhat
later than the transition time $T\approx 2.75M$ found above. Since in both
cases maximal slicing was used, the difference must come from a difference in
the initial slicing. In \cite{Libson94a} the \v{C}ade\v{z} lapse profile, which
is zero at the throats, was used initially, while in the current work the
initial lapse was one. Therefore in this case the evolution
proceeds somewhat faster until the lapse collapses, compared to the case with
the lapse initially collapsed at the throats, resulting in a shorter time for
the change of topology. It is not currently possible to use the \v{C}ade\v{z}
initial lapse profile in Cactus, so a direct comparison is unfortunately not
possible. However, the results are consistent.

\begin{figure}[ht]
\begin{center}
\includegraphics[width=75mm]{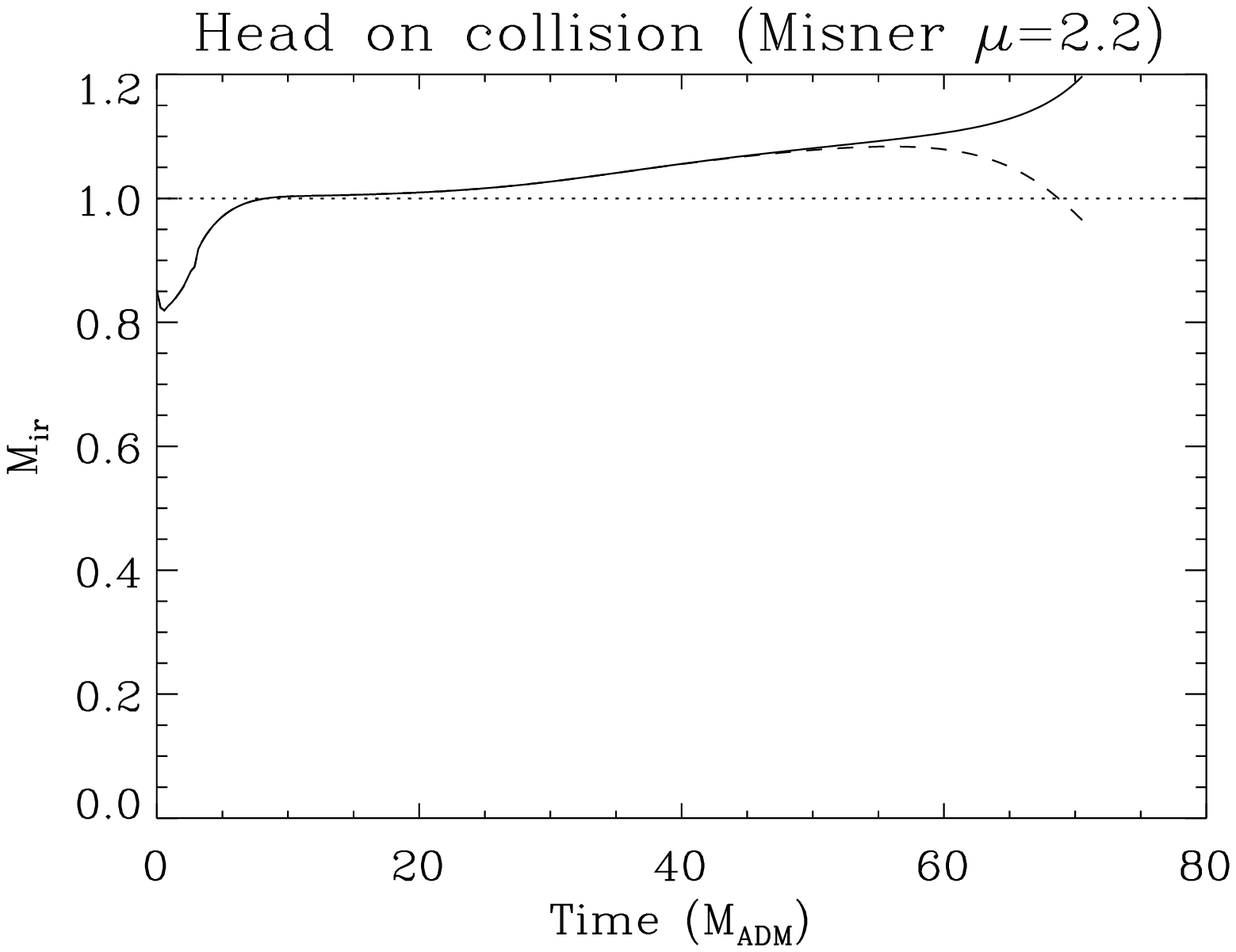}\hspace{2mm}
\includegraphics[width=75mm]{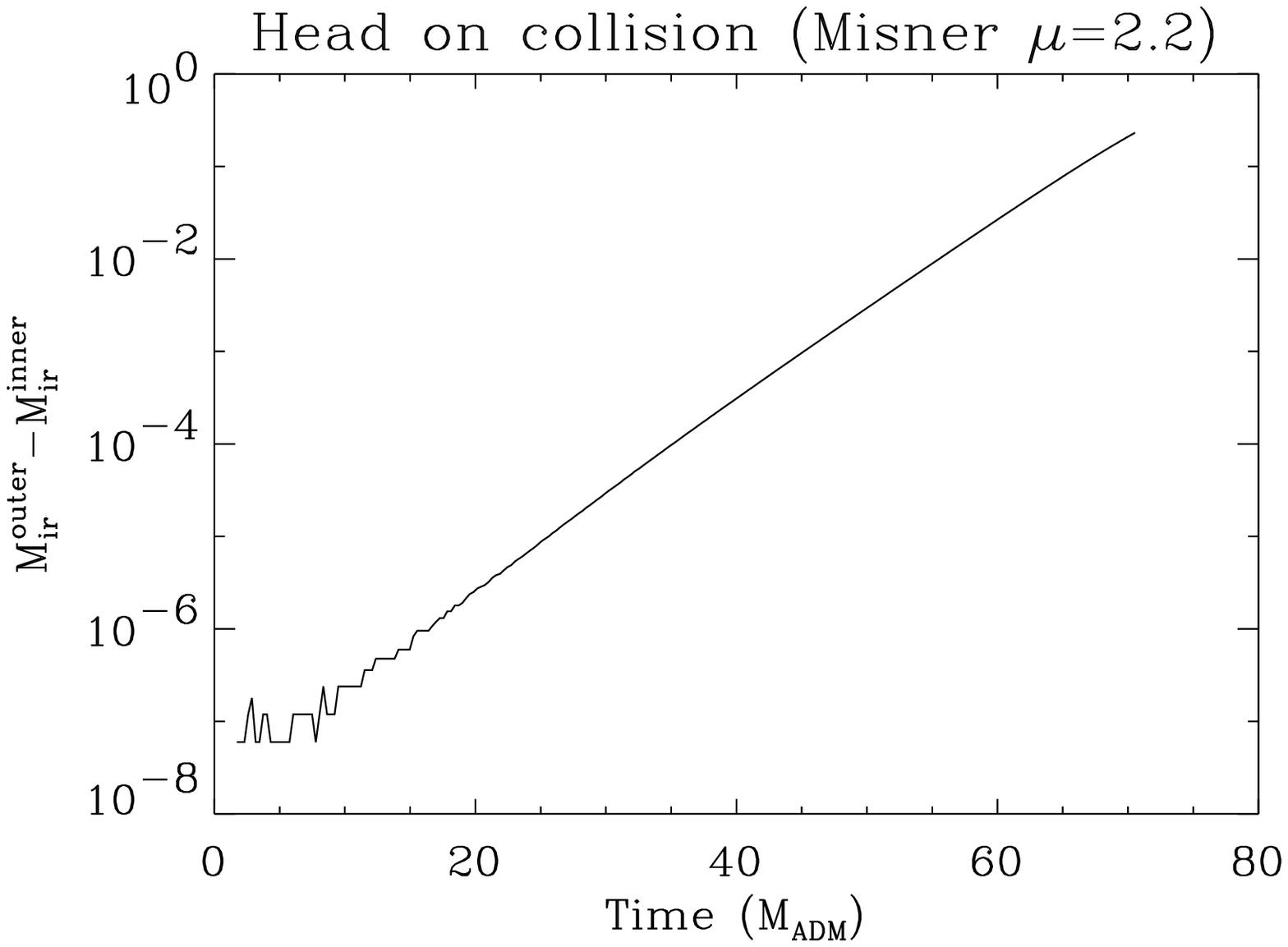}
\end{center}
\caption{The plot on the left shows the total irreducible mass of the 
         horizons in the Misner $\mu=2.2$ spacetime as a function of time.
         Shown is curves for two different choices of surfaces at $T=70M$. The
         solid line is for a sphere with radius $r_{O}=3.0M$ and the 
         dashed line is for a sphere with radius $r_{I}=2.5M$. The 
         horizontal dotted line denotes the ADM mass of the spacetime
         The plot on the left shows the difference in mass of the two surfaces.}
\label{fig:misner_mass}
\end{figure}

In the left plot in figure~\ref{fig:misner_mass} I show the total irreducible
mass of the horizons for the Misner $\mu=2.2$ spacetime as a function of
time for the two different choices for the initial surfaces mentioned
previously. The total irreducible mass is defined as
\[
M_{ir}=\sqrt{\frac{\sum_{i=1}^{n}{A_{i}}}{16\pi}},
\]
where $n$ is the number of horizons in the spacetime and $A_{i}$ is the area
of the $i$'th horizon.
As can be seen, even though $M_{ir}$ of the two surfaces are very different
at $T=70M$, they are almost indistinguishable on this plot for $T<50M$. The
fact that the curve crosses the dotted line and the subsequent rise in
$M_{ir}$ from about $T=20M$ to $T=50M$ are due to the lack of accuracy of the
underlying numerical data. The run was done at comparatively low resolution
and there are also reflections from the outer boundary. The initial dip in
$M_{ir}$ must be due to the low resolution (initially the horizons
are approximately $0.25M$ in radius) but as the horizons expand in
coordinates they become better resolved and the areas, and therefore $M_{ir}$,
correspondingly more accurate.

The right plot in figure~\ref{fig:misner_mass} shows the expected 
exponential convergence in the difference in $M_{ir}$ for the same two
choices of the initial surfaces. Looking closer it can be seen that the
convergence is in fact slightly better than exponential, but since the
horizon mass is growing slightly with time, this is to be expected. The
noise in the beginning, at $T<10M$, is a lot smaller than the truncation
error in either the interpolation and the finite differencing used in
obtaining the area.

\subsection{``Head on'' collision of 3 black holes}
\label{sec:3bh}
Since the Misner data is axisymmetric, the robustness of the code towards 
non-symmetric data was not really tested in the previous section. Therefore
I performed a numerical run with a three black hole spacetime. To keep things
simple I used Brill-Lindquist initial data \cite{Brill63} that contains no
linear and angular momentum. However, I placed the three black hole punctures
on a plane in coordinate space that was tilted with respect to the coordinate
axis and with slightly different coordinate distances between the black
holes. The masses were chosen to be equal $M_{1}=M_{2}=M_{3}=0.5$ giving a
system with an ADM mass of $M_{ADM}=1.5$\footnote{For Brill-Lindquist data
the ADM mass is just the sum of the individual black hole mass parameters,
while the bare masses are different from the mass parameters.}. The three
position vectors were $\mathbf{r}_{1}=(1.2,1.4,-0.8)$,
$\mathbf{r}_{2}=(-1.6,-0.6,-0.8)$ and $\mathbf{r}_{3}=(0.4,-0.8,1.6)$ giving
a system with the center of mass at the origin. The coordinate distances were
thus $d_{12}\approx 3.44$, $d_{13}\approx 3.35$ and $d_{23}\approx 3.13$. 

The spacetime was evolved similarly to the Misner spacetime in the previous
section, though a hyperbolic K-freezing slicing was used instead of maximal
slicing, simply because it is much faster. I used a fairly low resolution
of $\Delta = 0.2$, since the purpose was not to test the accuracy but rather
the robustness of the code. The run could then be performed on a workstation
with a moderate amount of disk requirements. Using the hyperbolic gamma 
freezing shift it was possible to evolve until $T=40=26.7M_{ADM}$ at which
time I stopped the evolution. The run could have continued a bit longer
but the Hamiltonian constraint was starting to grow
significantly\footnote{The growth of the Hamiltonian constraint was contained
mostly within the apparent horizon and was rather small in the exterior.} and
I estimated that the evolution time was long enough to locate the EH 
accurately enough for the initial $10M_{ADM}$ of evolution.

At about $T=10=6.7M_{ADM}$ a common AH appeared and at 
$T=40=26.7M_{ADM}$ it had evolved into an approximately spherical shape with 
a coordinate radius of $r_{AH}=3.6$. The deviations from sphericity was less 
than 1\%. Therefore it was safe to assume that a sphere of radius $r_{I}=3.5$
was completely contained within the EH at $T=40=26.7M_{ADM}$ and
this was chosen as the inner initial guess for the EH, while the
outer initial guess was chosen to be a sphere of radius $r_{O}=3.7$.

\begin{figure}[ht]
\begin{center}
\includegraphics[width=70mm]{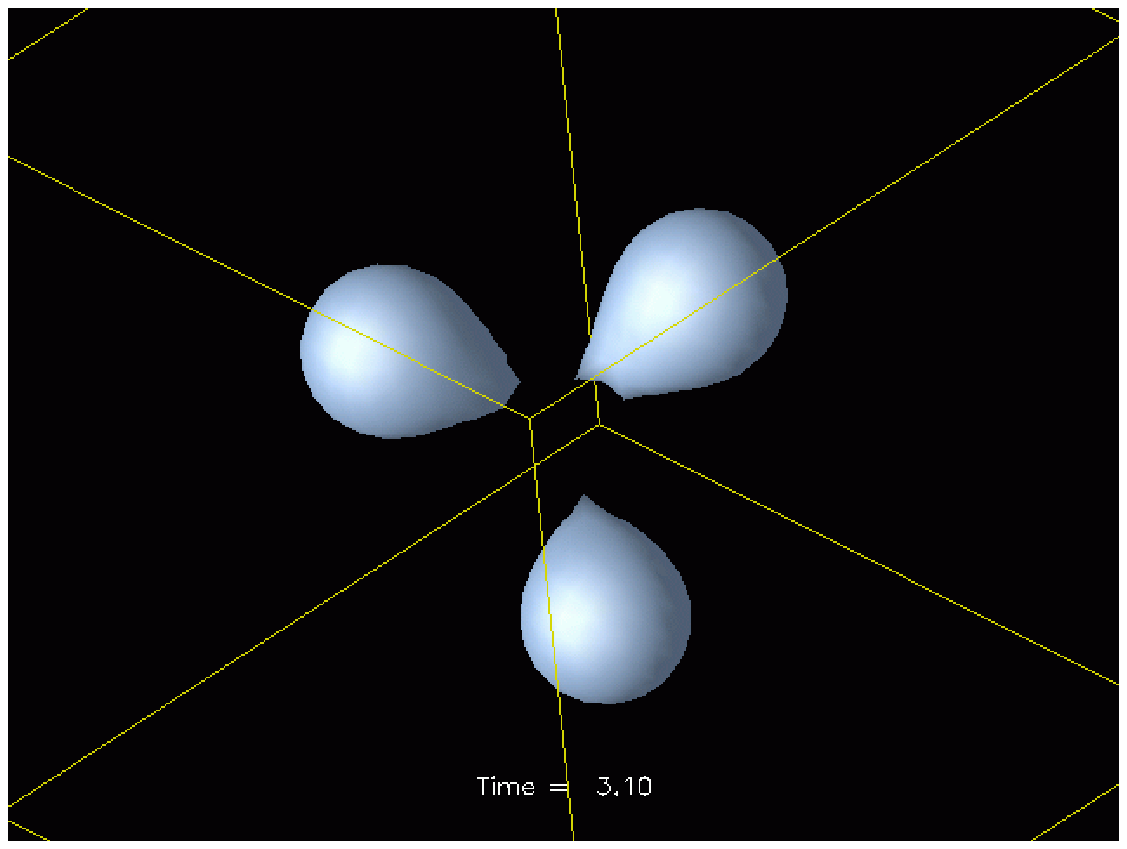}(a)
\includegraphics[width=70mm]{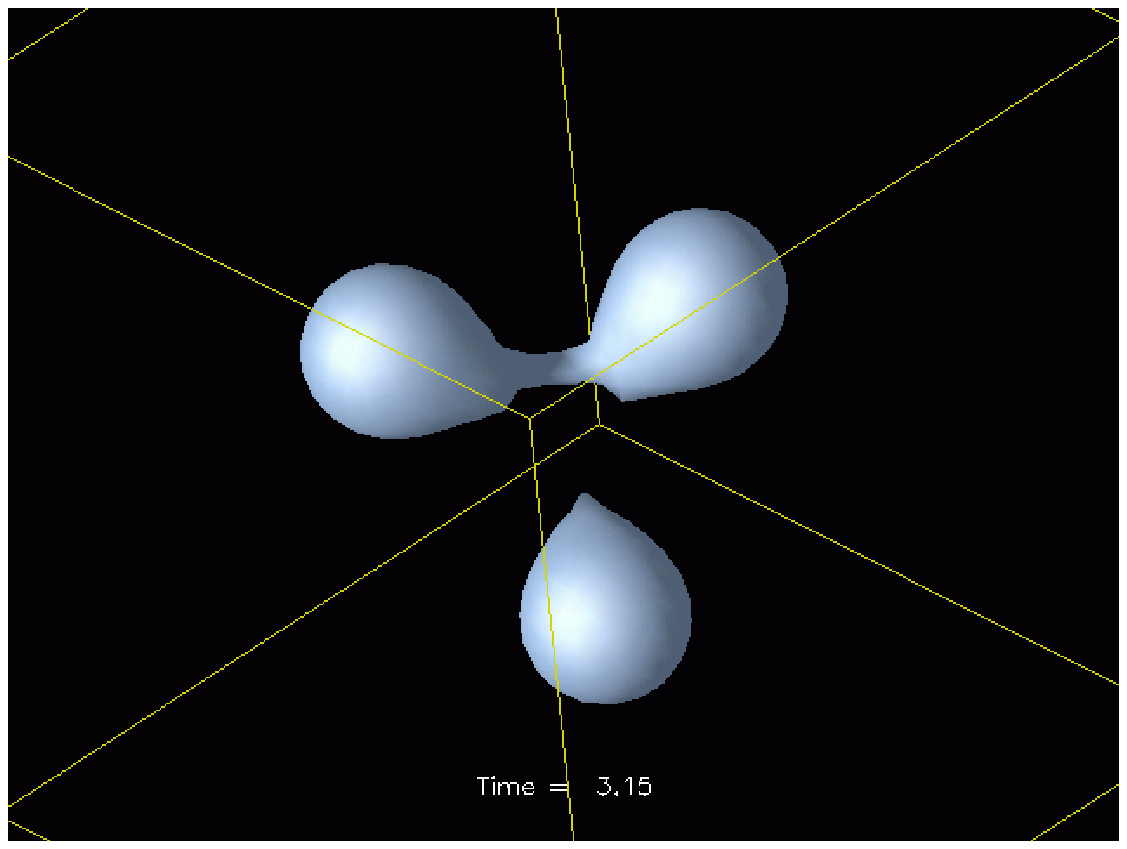}(b)\vspace*{2mm}
\includegraphics[width=70mm]{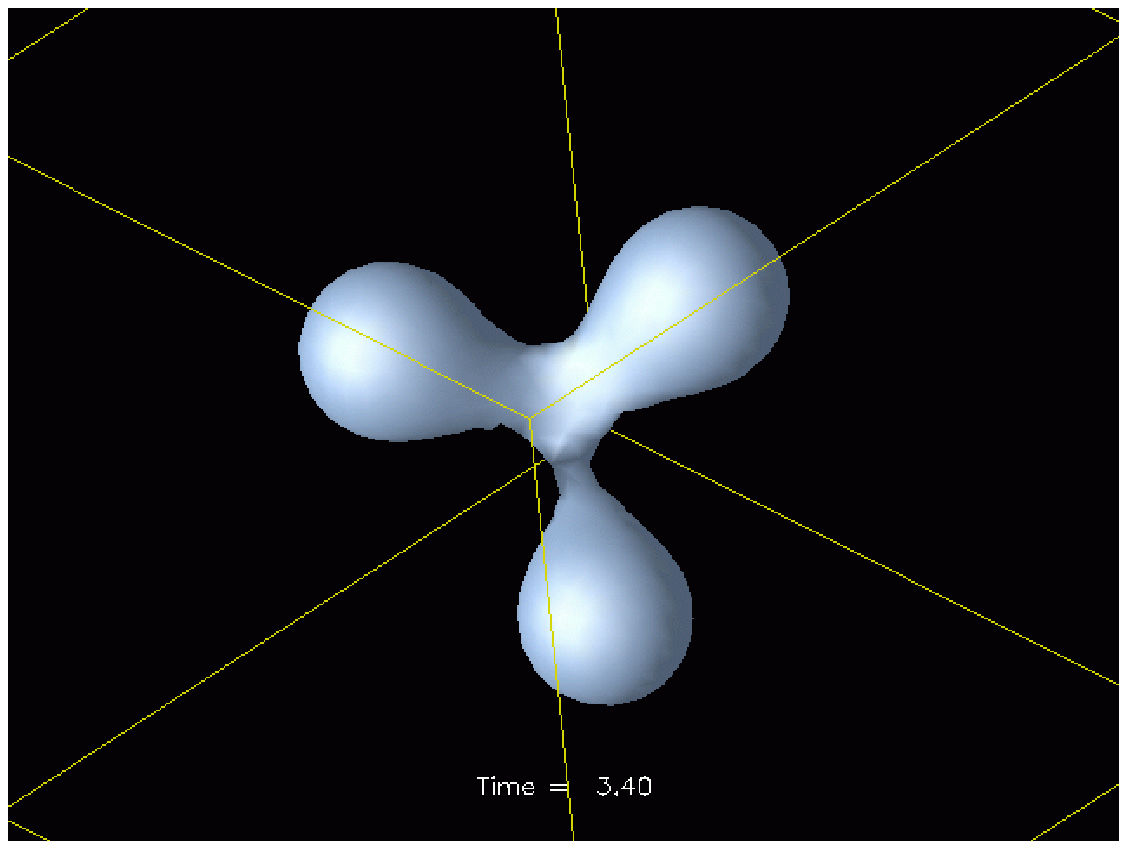}(c)
\includegraphics[width=70mm]{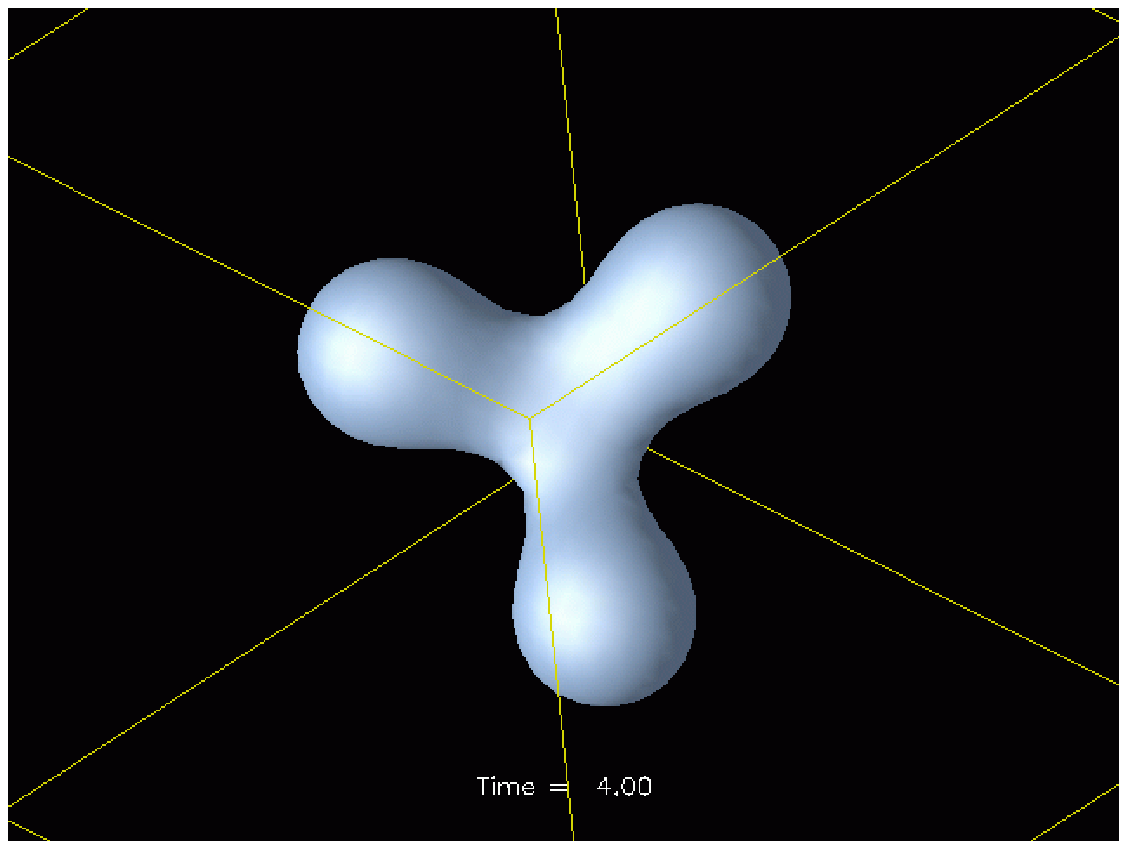}(d)
\end{center}
\caption{Four frames of the evolution of the EH in a 3 black hole 
         spacetime (Brill-Lindquist). Frame (a) is at $T=3.1=2.067M_{ADM}$, 
         frame (b) is at $T=3.15=2.1M_{ADM}$, Frame (c) is at 
         $T=3.4=2.267M_{ADM}$ and frame (d) is at $T=4.0=2.667M_{ADM}$. The
         straight lines in the plot, shows the bounding box of the numerical
         data. The outer boundary was much further out in the the evolution
         of the spacetime.}
\label{fig:3bh_evol}
\end{figure}

In figure~\ref{fig:3bh_evol} four frames showing the interesting part of the
evolution of the EH can be seen. 
Frame (a) shows 3 separated horizons at $T=2.067M_{ADM}$. The
individual horizons are clearly distorted and it can be seen that
they ``feel'' each other. Frame (b) shows that at $T=2.1M_{ADM}$
two of the horizons has merged while the third is still clearly well
separated. This has changed at $T=2.267M_{ADM}$ shown in frame (c) where the
final horizon has merged with the two other. In frame (d) the horizon is 
shown at $T=2.667M_{ADM}$, where it has started to increase in thickness. 
At later times the EH becomes more and more spherical.

The spatial and time resolution in the metric data is not high enough to
find instances of toroidal EHs as predicted in~\cite{Husa99a},
since these are generally short lived and have sharp features. However,
looking at frames (b) and (c), it does not take much imagination to envision
that a toroidal EH could appear for a short time, given higher spatial and
time resolution.

\begin{figure}[ht]
\begin{center}
\includegraphics[width=75mm]{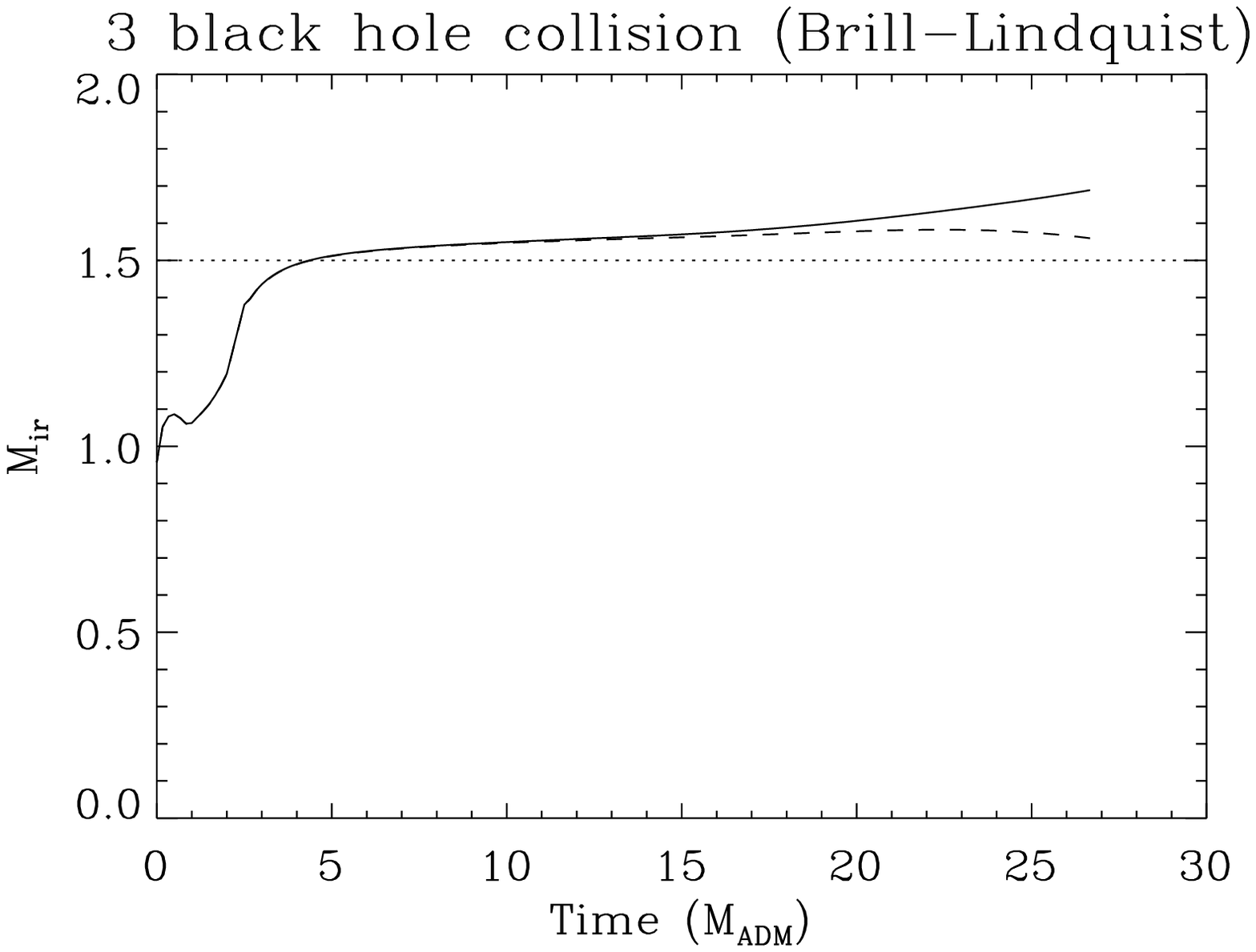}\hspace{2mm}
\includegraphics[width=75mm]{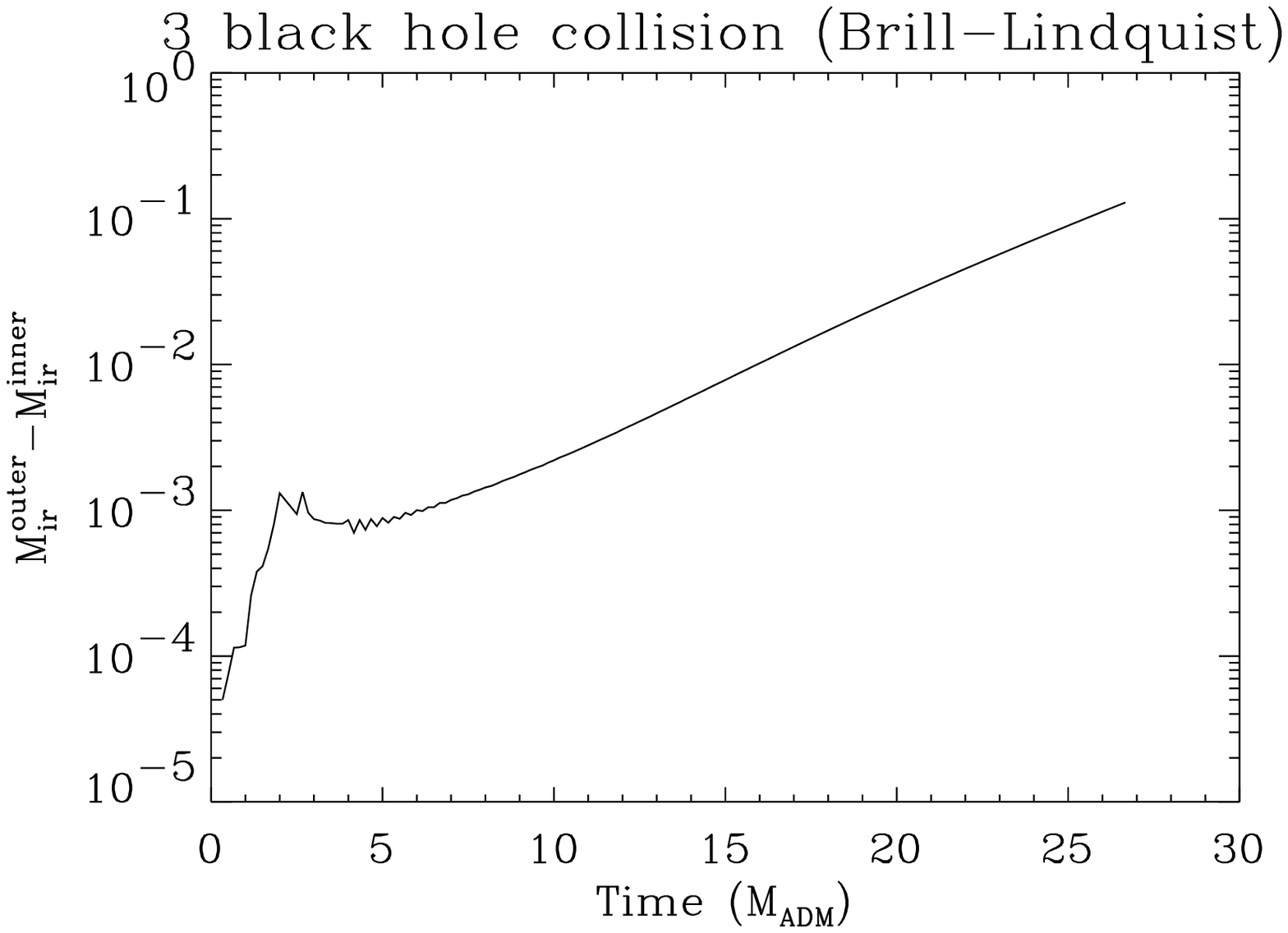}
\end{center}
\caption{The plot on the left shows $M_{ir}$ in the 
         3 black hole spacetime as a function of time. Shown is 
         curves for two different choices of surfaces at $T=27.7M_{ADM}$.
         The solid line is for a sphere with coordinate radius 
         $r_{O}=3.7$ and the dashed line is for a sphere with radius 
         $r_{I}=3.5$. The horizontal dotted line indicates the ADM mass,
         $M_{ADM}=1.5$, of the system. The plot on the left shows the
         difference in $M_{ir}$ for the two surfaces.}
\label{fig:3bh_area}
\end{figure}

The plots in figure~\ref{fig:3bh_area} shows the total irreducible mass of
the horizons, $M_{ir}$ in the spacetime as a function of time for the two
different choices of initial surfaces at $T=26.7M_{ADM}$. In these plots I
had to exclude a few points corresponding to times where the shape of the
horizon is so distorted that describing it with $r(\theta,\phi)$ around a
suitably chosen center is impossible (see e.g.\ frame (b) in 
figure~\ref{fig:3bh_evol}).
Again $M_{ir}$ goes beyond the limiting value set by the ADM mass of the
system, but in this case the resolution is even lower than in the Misner case
so this is not surprising. The convergence of the two surfaces are also here
nicely exponential, though the surfaces do not get as close together as in the
Misner case due to the shorter evolution time.

\section{Discussion}
\label{sec:discuss}
The EH finder presented here is a very general and accurate
method of locating and tracking EHs. It works both with
analytic and numerical metric data and on spacetimes with single or multiple
black holes and can handle changes of topology (and even though it has not
been demonstrated here, it should be able to handle toroidal EHs).
For analytic and stationary spacetimes the horizon can typically be located
to within a small fraction of a cell size (on the order of $1/10 \Delta$--
$1/100 \Delta$) at typical resolutions and the code shows second order
convergence with the grid resolution.

The currently implemented analysis tools are restricted to calculating the
area, centroid and polar and equatorial circumferences\footnote{The 
circumferences only makes sense in spacetimes with sufficient symmetries.} for
all the surfaces present in the data as long as the surfaces can be
represented in spherical coordinates as $r(\theta,\phi)$ around a point inside
the surface. As shown in figure~\ref{fig:3bh_evol} this is not always
possible, so a more general area integration routine is under consideration.

The robustness of the code was demonstrated by finding and tracking the
EH of a three black hole spacetime with no symmetries through
two changes of topology.

Analysis tools for calculating the Gaussian curvature of the horizon and for
tracking a congruence of individual generators of the horizon surface have been
shown in \cite{Masso98c} to be useful for extracting physical information
from highly dynamic spacetimes. With a congruence of horizon generators, it is
possible to construct the membrane paradigm quantities \cite{Thorne86}
and thus to study the shear and expansion of the horizon in detail. It should
be possible to track individual generators along with the horizon surface
using the information contained in the level set function. I am currently 
investigating a way to implement this powerful analysis tool.

The prediction of the generic formation of toroidal EHs (at least for a short
while) in non-symmetric spacetimes \cite{Husa99a} is interesting. I believe
that high resolution is required to see this numerically, but it might be
within the realm of possibility with this code.

This EH finder will be applied to a large number of different black
hole spacetimes that can be evolved long enough to reach an almost 
stationary state.

\ack
During the final phase of writing this paper, two other papers about EH
finding methods \cite{Caveny-Anderson-Matzner-2003a, Caveny-Matzner-2003a}
appeared on GR-QC.

Tests of the code have been performed on the Hitachi SR8000-F1 at the 
Leibniz-Rechenzentrum (LRZ), the IBM SP RS/6000 at the National Energy
Research Scientific Computing Center (NERSC), the IA-32 Linux cluser at
the National Center for Supercomputing Applications (NCSA), various 
workstations at the Albert-Einstein-Institut (AEI), my laptop and on the
new Linux cluster (Peyote) at the AEI. I would like to thank Ed Seidel,
Sascha Husa, Denis Pollney, Fransisco Guzman, Ian Hawke, Jonathan Thornburg
and Jeffrey Winicour for helpful discussions. On the practical side I would
like to thank Ian Hawke for implementing the MoL thorn in Cactus making it
easy to write an evolution code, Jonathan Thornburg for implementing the new
interpolator in Cactus that can return both the interpolated function and its
derivatives at the same time and Thomas Radke for helping me with the IO
aspects of the code. I would also like to thank the Tapir group at Caltech
for useful comments after my talk during my visit in February.

This work was supported by the EU Programme `Improving the Human
Research Potential and the Socio-Economic Knowledge Base' (Research Training
Network Contract HPRN-CT-2000-00137).
\appendix
\section{Level set evolution in Kerr-Schild coordinates}
\label{ap:kerrschild}
In this appendix I will show that, for a non-rotating black hole in
Kerr-Schild coordinates, an analytic solution to the evolution equation for
the level set function $f$ only exist for a limited time when evolving
backwards in time. I will only consider the spherically symmetric case 
$f=f(r)$. With the metric given by equation~\eref{eq:schw_ks} the relevant
components of the contravariant metric is
\[
g^{tt} = -\left ( 1+\frac{2M}{r} \right ), g^{tr} = \frac{2M}{r}, g^{rr} = 1-\frac{2M}{r},
\]
and the evolution equation for $f$ becomes
\begin{equation}
\frac{\partial f}{\partial t} = \frac{2M-r}{2M+r}\frac{\partial f}{\partial r}.
\end{equation}
For $r=2M$ this equation has the trivial solution
\[
f(t,r=2M)=c,
\]
where $c$ is a constant depending only on the initial value of $f$ at $r=2M$.
For $r\neq 2M$ this equation has the general solution
\begin{equation}
f(t,r) = h(u(t,r)) = h(t-r-4M\ln |2M-r|),
\end{equation}
where $u(t,r)=t-r-4M\ln |2M-r|$. I now limit myself to $r<2M$. At $t=0$ the 
following relation between $u(0,r)=u_{0}(r)$ and $r$ can be found
\[
u(0,r)=u_{0}(r)=-r-4M\ln (2M-r).
\]
The function $u_{0}(r)$ has the properties that 
$\lim_{r\rightarrow -\infty}=\infty$ and $\lim_{r\rightarrow 2M^{-}}=\infty$
and it has exactly one minimum in between at $r=-2M$ where the value is
$u_{0}(-2M)=2M-4M\ln(4M)$. The range of $u(0,r)=u_{0}(r)$ is therefore the
unbounded interval $[2M-4M\ln(4M),\infty)$, which again is the domain of 
the $h(u)$. The function, $h(u)$, does not change in time, so in order
for the solution, $f(t,r)$ to exist, $u(t,r)$ must satisfy 
$u(t,r)\geq 2M-4M\ln(4M)$.
From this it can be seen that a solution exists only at $r=0$ for
\[
t\geq 2M-4M\ln 2\approx -0.77259 M.
\]
There is no limit to the existence of a solution when evolving forward in time.
In the solution, the problem shows up as the development of infinite
derivatives, $\partial f(t,r)/\partial t$ and $\partial f(t,r)/\partial r$,
at $r=0$ in a finite time. Excising a region around $r=0$ does not help, as
the point of infinite gradient moves outward and will reach the boundary of
the excision region in a finite time. The time for the development of infinite
gradients is independent of the initial shape of $f(0,r)$ and are purely a
property of the metric. A non rotating black hole in isotropic coordinates
do not have the same property. In that case it takes infinite time to develop
infinite gradients.
\section{Second order upwinded derivatives}
\label{ap:upwind}
Define $a_{l}$ and $a_{r}$ as the second order one sided derivatives of $f$ in
the $x$-direction as
\begin{eqnarray}
a_{l} & = & \frac{3f_{i,j,k}-4f_{i-1,j,k}+f_{i-2,j,k}}{2\Delta}, \nonumber \\
a_{r} & = & \frac{-3f_{i,j,k}+4f_{i+1,j,k}-f_{i+2,j,k}}{2\Delta}. \nonumber
\end{eqnarray}
Then define the negative and positive parts of $a_{l}$ and $a_{r}$ as
\begin{eqnarray}
a_{l}^{-} & = \frac{a_{l}-\|a_{l}\|}{2}, \mbox{ }
a_{l}^{+} & = \frac{a_{l}+\|a_{l}\|}{2}, \nonumber \\
a_{r}^{-} & = \frac{a_{r}-\|a_{r}\|}{2}, \mbox{ }
a_{r}^{+} & = \frac{a_{r}+\|a_{r}\|}{2}. \nonumber
\end{eqnarray}
If $f_{i}\geq 0$ then
\[
\frac{\partial f}{\partial x}\approx\left \{\begin{array}{lcr}
a_{l}^{+} & \mbox{if} & a_{l}^{+} > -a_{r}^{-} \\
a_{r}^{-} & \mbox{otherwise}. &
\end{array} \right .
\]
If $f_{i}<0$ then
\[
\frac{\partial f}{\partial x}\approx\left \{\begin{array}{lcr}
a_{l}^{-} & \mbox{if} & -a_{l}^{-} > a_{r}^{+} \\
a_{r}^{+} & \mbox{otherwise}. &
\end{array} \right .
\]
Derivatives with respect to $y$ and $z$ are approximated by corresponding
equations. This method of upwinding derivatives is a generalization of the
first order method presented in \cite{Sussman94} to second order derivatives.
The first order method is shown to be a consistent monotone scheme for the
equation under consideration in \cite{Sussman94}. The above second order
generalization is not guaranteed to be a consistent monotone scheme for the
equations considered here, but it works nicely in practice.

\section*{References}
\bibliographystyle{prsty}
\bibliography{bibtex/references}

\begin{thebibliography}{10}

\bibitem{Alcubierre02a}
M. Alcubierre {\it et~al.}, Phys. Rev. D {\bf 67},  084023  (2003).

\bibitem{Hughes94a}
S. Hughes {\it et~al.}, Phys. Rev. D {\bf 49},  4004  (1994).

\bibitem{Anninos94f}
P. Anninos {\it et~al.}, Phys. Rev. Lett. {\bf 74},  630  (1995).

\bibitem{Libson94a}
J. Libson {\it et~al.}, Phys. Rev. D {\bf 53},  4335  (1996).

\bibitem{Masso98c}
J. Mass{\'o}, E. Seidel, W.-M. Suen, and P. Walker, Phys. Rev. D {\bf 59},
  064015  (1999).

\bibitem{Nakamura84}
T. Nakamura, Y. Kojima, and K. Oohara, Phys. Lett. {\bf 106A},  235  (1984).

\bibitem{Tod91}
K.~P. Tod, Class. Quantum Grav. {\bf 8},  L115  (1991).

\bibitem{Kemball91a}
A.~J. Kemball and N.~T. Bishop, Class. Quantum Grav. {\bf 8},  1361  (1991).

\bibitem{Baumgarte96}
T.~W. Baumgarte {\it et~al.}, Phys. Rev. D {\bf 54},  4849  (1996).

\bibitem{Thornburg95}
J. Thornburg, Phys. Rev. D {\bf 54},  4899  (1996).

\bibitem{Gundlach97a}
C. Gundlach, Phys. Rev. D {\bf 57},  863  (1998), gr-qc/9707050.

\bibitem{Libson94b}
P. Anninos {\it et~al.}, Phys. Rev. D {\bf 58},  024003  (1998).

\bibitem{Alcubierre98b}
M. Alcubierre {\it et~al.}, Class. Quantum Grav. {\bf 17},  2159  (2000).

\bibitem{Shoemaker-Huq-Matzner-2000}
D.~M. Shoemaker, M.~F. Huq, and R.~A. Matzner, Phys. Rev. D {\bf 62},  124005
  (12~pages)  (2000).

\bibitem{Huq00}
M.~F. Huq, M.~W. Choptuik, and R.~A. Matzner, Phys. Rev. D {\bf 66},  084024
  (20002), gr-qc/0002076.

\bibitem{Schnetter02a}
E. Schnetter,   (2002), gr-qc/0206003.

\bibitem{Thornburg2003:AH-finder}
J. Thornburg,   (2003), in preparation.

\bibitem{Sussman94}
M. Sussman, P. Smereka, and S. Osher, J. Comp. Phys. {\bf 114},  146  (1994).

\bibitem{Allen99a}
G. Allen, T. Goodale, and E. Seidel,  in {\em 7th Symposium on the Frontiers of
  Massively Parallel Computation-Frontiers 99} (IEEE, New York, 1999).

\bibitem{Goodale02a}
T. Goodale {\it et~al.},  in {\em Vector and Parallel Processing - VECPAR'2002,
  5th International Conference, Lecture Notes in Computer Science} (Springer,
  Berlin, 2003).

\bibitem{Goodale03c}
T. Goodale,   (2003), in preparation.

\bibitem{Misner60}
C. Misner, Phys. Rev. D {\bf 118},  1110  (1960).

\bibitem{Anninos96c}
P. Anninos, J. Mass{\'o}, E. Seidel, and W.-M. Suen, Physics World {\bf 9},  43
   (1996).

\bibitem{Shibata95}
M. Shibata and T. Nakamura, Phys. Rev. D {\bf 52},  5428  (1995).

\bibitem{Baumgarte99}
T.~W. Baumgarte and S.~L. Shapiro, Physical Review D {\bf 59},  024007  (1999).

\bibitem{Alcubierre99c}
M. Alcubierre, B. Br\"{u}gmann, M. Miller, and W.-M. Suen, Phys. Rev. D {\bf
  60},  064017  (1999).

\bibitem{SeidelPrivateComm}
E. Seidel, private communication.

\bibitem{Brill63}
D. {B}rill and R. Lindquist, Phys. Rev. {\bf 131},  471  (1963).

\bibitem{Husa99a}
S. Husa and J. Winicour, Physical Review D {\bf 60},  084019  (1999).

\bibitem{Thorne86}
{\em Black Holes: The Membrane Paradigm}, edited by K.~S. Thorne, R.~H. Price,
  and D.~A. Macdonald (Yale University Press, London, 1986).

\bibitem{Caveny-Anderson-Matzner-2003a}
S.~A. Caveny, M. Anderson, and R.~A. Matzner,   (2003), preprint gr-qc/0303099.

\bibitem{Caveny-Matzner-2003a}
S.~A. Caveny and R.~A. Matzner,   (2003), preprint gr-qc/0303109.

\end{thebibliography}

\end{document}